\documentclass{PoS}

\usepackage{amsmath}
\usepackage{amsthm}
\usepackage{graphicx}

\title{Dual simulation of finite density\\ lattice QED at large mass\footnote{
This project is partly supported by DFG TR55, ''Hadron Properties from Lattice QCD'' 
and by the Austrian Science \hspace*{7mm} Fund FWF Grant. Nr. I 1452-N27.}}

\ShortTitle{Dual simulation of finite density lattice QED at large mass}

\author{Michael Kniely\\
        Institut f\"ur Mathematik und Wissenschaftliches Rechnen, University of Graz, 8010 Graz, Austria\\
        E-mail: \email{michael.kniely@uni-graz.at}}
        
\author{\speaker{Christof Gattringer}\\
        Institut f\"ur Physik, University of Graz, 8010 Graz, Austria\\
        E-mail: \email{christof.gattringer@uni-graz.at}}

\abstract{We discuss a mapping of lattice QED with two flavors and a chemical potential
to dual variables, which are surfaces for the gauge fields and loops for the fermions. 
The gauge fields are completely dualized and the corresponding dual variables are integer valued plaquette occupation numbers with constraints
that lead to a structure of surfaces that are either closed or bounded by fermion loops. The fermion loops are obtained from a resummed 
hopping expansion (large mass expansion) 
of the determinant of the Wilson-Dirac operator. The loops can come with both positive and negative signs. 
We identify a sub-class of loops, which we refer to as quasi-planar loops, where the total sign is positive. For this sub-class a dual Monte Carlo 
simulation is possible and we discuss its implementation and some results. In particular we address condensation phenomena at finite 
chemical potential.}

\FullConference{The 32nd International Symposium on Lattice Field Theory\\
		 23-28 June, 2014\\
		 Columbia University New York, NY}

\begin{document}

\section{Motivation}
The Monte Carlo simulation of lattice field theories at finite chemical potential $\mu$ is a notoriously difficult task due to the complex action 
problem: At nonzero $\mu$ the action $S$ is complex and the Boltzmann factor $e^{-S}$ cannot be used as a probability weight. A powerful
approach to overcome the complex action problem is to find a re-parametrization of the partition sum $Z$ in terms of new variables, such 
that $Z$ is a sum over only real and positive contributions. Then a Monte Carlo simulation can be set up in terms of these so-called
dual variables, which are loops for matter fields and surfaces for gauge fields. The dual approach has been successfully implemented for 
several interesting lattice field theories at finite density (see, e.g., \cite{dual1,dual2,dualu1} for examples). 

\hspace*{10mm}
So far the dualization was mainly restricted to bosonic theories, since the fermions contribute additional minus signs for the matter loops, which
come from the Pauli principle and the traces over the $\gamma$-matrices. However, integrating out the gauge fields leads to additional 
constraints for admissible dual configurations of loops and surfaces, which can lead to cancellations of the fermion 
signs. In order to study these emerging
dual structures we here analyze lattice QED with two flavors of Wilson fermions and a chemical potential. For this model the gauge fields can be
dualized completely (this is possible for all abelian gauge theories) and one can focus on studying the structure of the fermion loops interacting
with the gauge surfaces. In this contribution we study the loops as obtained from a partially summed hopping expansion and their interaction
with the dual gauge surfaces. We show that for certain classes of loops (quasi-planar loops), the total signs are positive and we present results
for a dual simulation at finite chemical potential.
 
\section{Definition of the model and hopping expansion of the fermion determinant}
In the conventional representation the partition sum of lattice QED with two flavors is given by
\begin{equation}
Z = \int \mathcal{D}[U] \; e^{-S_G[U]} \, \det D[U] \, \det D[U^*] \; .
\end{equation}
Here the fermions are already integrated out and only the path integral over the link variables  $U_\nu(n) \in U(1)$ remains, where 
$\mathcal{D}[U]$ denotes 
the product of the Haar measures for all gauge links.
For the gauge action we use the Wilson form, $S_G[U] = - \beta \sum_p \; \mbox{Re} \; U_p$, where $\beta$ denotes the inverse gauge 
coupling, the sum runs over all plaquettes $p$, and $U_p$ is the product of link variables along the contour of $p$. We consider two 
mass-degenerate flavors of opposite charge, such that Gauss' law is obeyed. As a consequence the conjugate gauge links $U^*_\nu(n)$ are used 
in the second fermion determinant. The Wilson Dirac operator is denoted by $D[U] = 1 - \kappa H[U]$, with the standard Wilson hopping matrix
$H[U](n,m) = \sum_{\nu} \Big[ e^{\mu\delta_{\nu,4}} \, \frac{ 1 - \gamma_\nu}{2} \, U_\nu(n) \, \delta_{n+\hat\nu,m}
\; + \;  e^{- \mu\delta_{\nu,4}} \, \frac{ 1 + \gamma_\nu}{2} \, U_\nu(n - \hat\nu)^* \, \delta_{n-\hat\nu,m} \Big]$, where
 $\kappa$ is the hopping parameter and $\mu$ the chemical potential. Both parameters are chosen to have the same values  
 for the two flavors (but remember that for the second flavor the complex conjugate link variables are used in the hopping matrix). 
 $n$ and $m$ label the sites of a $N_S^3 \times N_T$ lattice with
boundary conditions that are periodic in the three spatial directions and anti-periodic in time. 
 
 \hspace*{10mm}
In order to obtain the fermion loops of the dual representation we apply hopping expansion of the determinant (see, e.g., \cite{book} for an 
elementary introduction), i.e., an expansion in small $\kappa$ (large mass).
Using the trace-log formula the fermion determinant can be rewritten as $\det D = \exp( - \sum_n \frac{\kappa^n}{n} \; \mbox{Tr} \; H^n )$,
where we have already expanded in $\kappa$ the logarithm in the exponent. The traces $\mbox{Tr} \; H^n$ over powers of the
hopping matrix give rise to non-backtracking closed loops of length $n$ which are dressed with the corresponding hopping terms
along their contour. Thus the 
$ \sum_n \frac{\kappa^n}{n} \; \mbox{Tr} \; H^n $ can be viewed as a sum over all closed loops. Among the set of all loops there are also 
those where a simple loop is iterated several times. Thus we can split the sum into a sum over all non-iterating loops and a sum over the
iterations of the loops. The sum over the non-iterating loops in the exponent we write as a product over the 
exponentials, and we keep in the exponent 
only the sum over the iterations. However, this sum over iterations in the exponent can be written again as a determinant over only the  
Dirac indices of the product of hopping terms along the contour of the loop. Thus we have brought the determinant into the form
\begin{equation}
\det D = \prod_{l \in \mathcal{L}_\mathrm{NI}} \det \left( 1 - c_l \Gamma_l \right) = 
\prod_{l \in \mathcal{L}_\mathrm{NI}} \chi_{c_l\Gamma_l}(1) = \prod_{l \in \mathcal{L}_\mathrm{NI}} 
\left(1 - \mbox{tr} \, (\Gamma_l) \, c_l + \mbox{sumdet}(\Gamma_l) \, c_l^2 \right) \; .
\label{detprod}
\end{equation}
Here $\mathcal{L}_\mathrm{NI}$ is the set of all oriented, non-backtracking, non-iterating loops $l$, $\Gamma_l$ is the product of the matrices
$(1 \mp \gamma_\nu)/2$ along the loop $l$, and $c_l$ is given by 
\begin{equation}
c_l \; = \; (-1)^{W(l)} \, e^{\mu N_T W(l)} \, \kappa^{|l|} \, U_l \; .
\label{coeffs}
\end{equation}
By $W(l)$ we denote 
the number of times the loop $l$ winds around compact time (here we obtain an extra minus sign from every winding due to the 
anti-periodic temporal boundary conditions of the fermions), and $U_l$ is the product of the link variables along the loop $l$. 

\hspace*{10mm}
The determinant in (\ref{detprod}) now has the form of a product over loops in $\mathcal{L}_\mathrm{NI}$, where each loop $l$ can contribute 
in three ways: The first possibility is a factor of $1$, which is what we refer to as {\sl ''not activated''}. However, the loop $l$ can also contribute
a factor of $- \mbox{tr} \, (\Gamma_l) \, c_l$ ({\sl "single occupancy"}), or finally, $l$ contributes a factor $\mbox{sumdet}(\Gamma_l) \, c_l^2$,
which we refer to as  {\sl "double occupancy"} of the loop $l$.

\hspace*{10mm}
In the second step
of (\ref{detprod}) we have written the determinants for the individual loops $l$ with the characteristic polynomial 
$\chi_{c_l\Gamma_l}(\lambda) = \det(\lambda - c_l\Gamma_l)$ at $\lambda = 1$. 
A characteristic polynomial can be expanded in a sum over determinants of reduced matrices (some columns and rows deleted). Since the 
$\Gamma_l$ are products of the rank-2 projectors $(1 \mp \gamma_\nu)/2$, only terms where two or three columns and rows are deleted remain
in that expansion. This leads to the rhs.\ of (\ref{detprod}), where we introduced  
$\mbox{sumdet} \, (A) = \sum_{<n,m>} \det A_{<n,m>}$, where the sum runs over all ordered pairs  $<n,m>$ with $1 \leq n < m \leq 4$, and 
$A_{<n,m>}$ is the reduced matrix where the $n$-th and $m$-th rows and columns are removed. 

\hspace*{10mm}
From (\ref{detprod}) and the form of the $c_l$  
it is obvious that only temporally winding loops couple  to the chemical potential. Thus we now restrict the 
sum over loops in (\ref{detprod}) to winding loops, where for now we do not distinguish between winding around compact time or 
space directions.
We consider so-called quasi-planar loops, which are loops that after one or more steps in a direction perpendicular to the winding direction again 
continue in the winding direction.  A key step is the evaluation of
the signs coming from the Clifford algebra, which in our presentation corresponds to computing $\mbox{tr} \, (\Gamma_l)$ and 
$\mbox{sumdet}(\Gamma_l)$. It has been known for a long time that for planar loops $l$ the traces over the $\Gamma_l$ 
can be computed in closed form \cite{nucu}. Using similar techniques as in \cite{nucu} one finds for quasi-planar loops 
\begin{equation}
   \mbox{tr} \, (\Gamma_l) = \left( \frac{1}{2} \right)^{t-1} \mbox{\quad and \quad} \mbox{sumdet} \, (\Gamma_l) = \left( \frac{1}{2} \right)^{2t} \; ,
   \label{loopshape}
\end{equation}
where $t$ denotes the number of deviations from the main direction. 

\section{Integrating out the gauge fields and final form of the dual representation}
The second step in the dualization is to integrate out the gauge fields. They come from two sources: 
The fermion loops are dressed with gauge link variables along their contour and in the plaquettes of the gauge action we have products of four 
link variables along all unit squares of the lattice. The important key formula for integrating the link variables is (the $j_\nu(n)$ are arbitrary integers)
\begin{equation}
\int \! \mathcal{D}[U] \; \prod_{n,\nu} U_\nu(n)^{\, j_\nu(n)} \; = \; \prod_{n,\nu} \delta_{j_\nu(n),0} \; , 
\label{gaugedelta}
\end{equation}
i.e., only contributions where the links
along the fermion loops are compensated with links from the plaquettes of the gauge action do survive
(see also the discussion in \cite{dualu1}). To organize the gauge integration
we expand the Boltzmann factor of the gauge action using the well known series
\begin{equation}
e^{-S_G[U]} \; = \; \prod_p  e^{ \frac{\beta}{2} \, ( U_p + U_p^* ) }  \; = \;  \prod_p \sum_{n_p = -\infty}^\infty  \, 
I_{|n_p|}\!\left( \beta /2 \, \right)  \; U_p^{ \,n_p} \; ,
\label{gaugebessel}
\end{equation}
where $I_{|n_p|}\!\left( \beta /2 \   \right)$ 
denotes the modified Bessel functions, and the integers $n_p$ are referred to as the plaquette occupation numbers for the plaquette $p$. These 
plaquette occupation numbers describe the dual degrees of the U(1) gauge fields. The series in (\ref{gaugebessel}) has terms with
arbitrary integer powers $n_p$ of a plaquette $U_p$, where negative powers correspond to an orientation 
of the plaquette in the mathematically negative sense. Only those configurations of the plaquette occupation numbers $n_p$ survive, 
where the occupied plaquettes compensate the links along the fermion loops.  
This implies, that for obtaining non-vanishing dual configurations one builds up surfaces of occupied plaquettes $p$ (i.e., $n_p \neq 0$), and
the surfaces are either closed or bounded by the fermion loops. Note that for the second flavor the link variables along the
loops are complex conjugate and plaquettes with the opposite orientation have to be used to compensate that flux.
  
\hspace*{10mm}
With (\ref{gaugedelta}), (\ref{gaugebessel}) and the results for the loops from the previous section we obtain 
\begin{equation}
Z \; = \;  \sum_{\{l_1, l_2, n\}} \frac{e^{\, \mu(W_1 + W_2)} \; \kappa^{ \, L_1 + L_2}}{ 2^{T_1 + T_2}} \; \prod_p I_{|n_p|}\!\left( \beta /2 \, \right)  \; .
\label{zfinal}
\end{equation}
The sum is over the sets $\{l_1\}, \{l_2\}$ of all quasi-planar, non-iterating loops $l_1$ and $l_2$ representing the dual fermion degrees of freedom
for the two flavors, where each loop can have either single or double occupation. By 
$W_1, W_2$ we denote the corresponding total winding numbers for all loops of the two flavors. $L_1$ and $L_2$ denote
the total length of all the loops of the two flavors, and $T_1$ and $T_2$ are the shape factors of the loops, computed from the number of
deviations of the loops from the winding 
direction according to (\ref{loopshape}), with the trace being used for single occupation and sumdet for double 
occupation. In $Z$ we also sum over the dual degrees of freedom for the gauge fields: We sum over all 
admissible configurations $\{ n \}$ of the plaquette occupation number $n_p$, where ''admissible'' refers to the fact that the total flux along the 
fermion loops is compensated by plaquettes. As mentioned before, this amounts to summing over surfaces made from occupied plaquettes,
where the surfaces are either closed or bounded by fermion loops (see also the discussion in \cite{dualu1}).

\hspace*{10mm}
An important constraint comes from Gauss' law which requires an equal number of loops that have their
links oriented forward and backward in time (double occupied loops count twice). In the dual language this is manifest in the fact that only for 
vanishing link flux from the loops we can obey all constraints with finitely many occupied plaquettes. These constraints can be obeyed for example 
by a forward and a backward winding fermion loop of the same flavor, or by two loops of different flavor but with the same orientation. In all cases 
the total number of occupied loops is even (double occupation counts twice) and we do not have any remaining minus  
signs in (\ref{zfinal}) from single occupation of loops or the anti-periodic temporal boundary conditions of the fermions.

\section{Dual simulation of the model and numerical results}

We consider the model described by (\ref{zfinal}) in two variants: The first variant uses the full dynamics of the  
quasi-planar loops, while in the second variant we restrict the dual fermionic degrees further and only consider linear (straight) loops winding 
in time direction. The latter case corresponds to static charges and we refer to it as the "linear loop model". 

\hspace*{10mm}
Both variants of the model are updated with local Monte Carlo steps. For the dual gauge degrees of freedom we use two steps that can update also 
pure gauge theory: 1) We propose a change of all plaquette occupation numbers $n_p$
of an elementary 3-cube by $\pm 1$, arranged such that the net flux 
on all edges of the cube remains unchanged. This proposal is accepted with a Metropolis step according to the weights in (\ref{zfinal}) 
and leads to an admissible configuration. 2) We propose a change of all plaquette  occupation numbers $n_p$ on one of the 6 
coordinate planes by $\pm 1$. This again leads to an admissible configuration which we 
accept with a Metropolis step.

\hspace*{10mm}
For the linear loop model two updates of the fermion loops are used: 1) We insert pairs of temporally winding linear loops at the same spatial 
position, such that the constraints are kept intact. This can
either be a pair of oppositely oriented loops of the same flavor, or two loops with the same orientation but with different flavor. In the Metropolis 
acceptance step of the latter, also the chemical potential enters. 2) We allow a linear loop to hop in space. In that case we need to change the
plaquette occupation numbers $n_p$ in a strip that connects the old and the new position of the loop, and the corresponding weights determine the 
Metropolis acceptance.
Finally, for updating the full quasi-planar model, we also consider local, quasi-planar deformations of the loop. Here the Metropolis acceptance 
is computed from the changing length and shape of the loop, as well as from the weights of the plaquette occupation numbers that have to 
be changed in order to remain in the set of admissible configurations. 

\hspace*{10mm}
Various bulk observables were considered in our study -- they can be simply obtained as derivatives with respect to the parameters 
$\beta$, $\kappa$ and $\mu$ applied to the conventional and the dual representations. Here we restrict ourselves to presenting results for the
particle number density $n = 1/N_S^3 N_T \partial  \ln Z / \partial \mu $, which we use for studying condensation phenomena.

\begin{figure}[t]
	\centering
	\includegraphics[width=0.49\textwidth,clip]{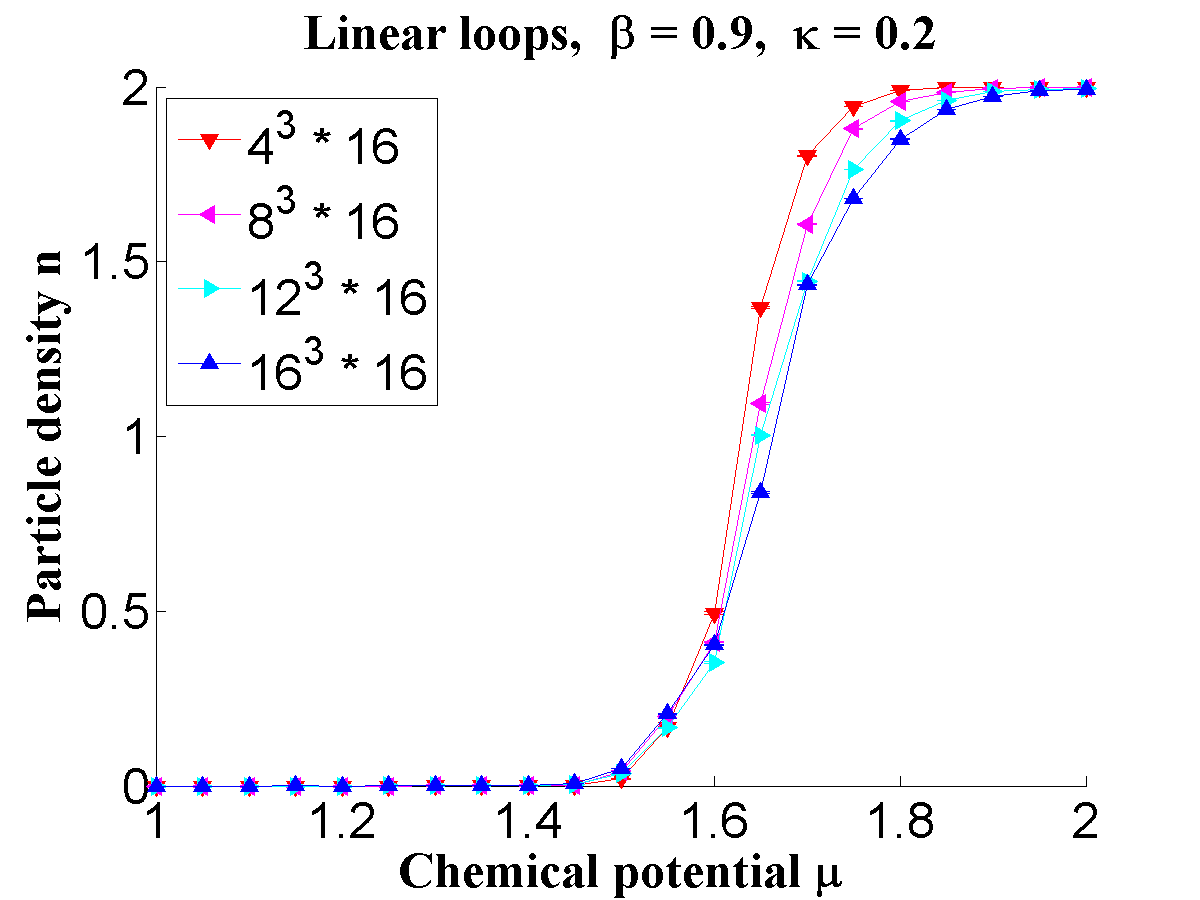}
	\hspace{0.005\textwidth}
	\includegraphics[width=0.49\textwidth,clip]{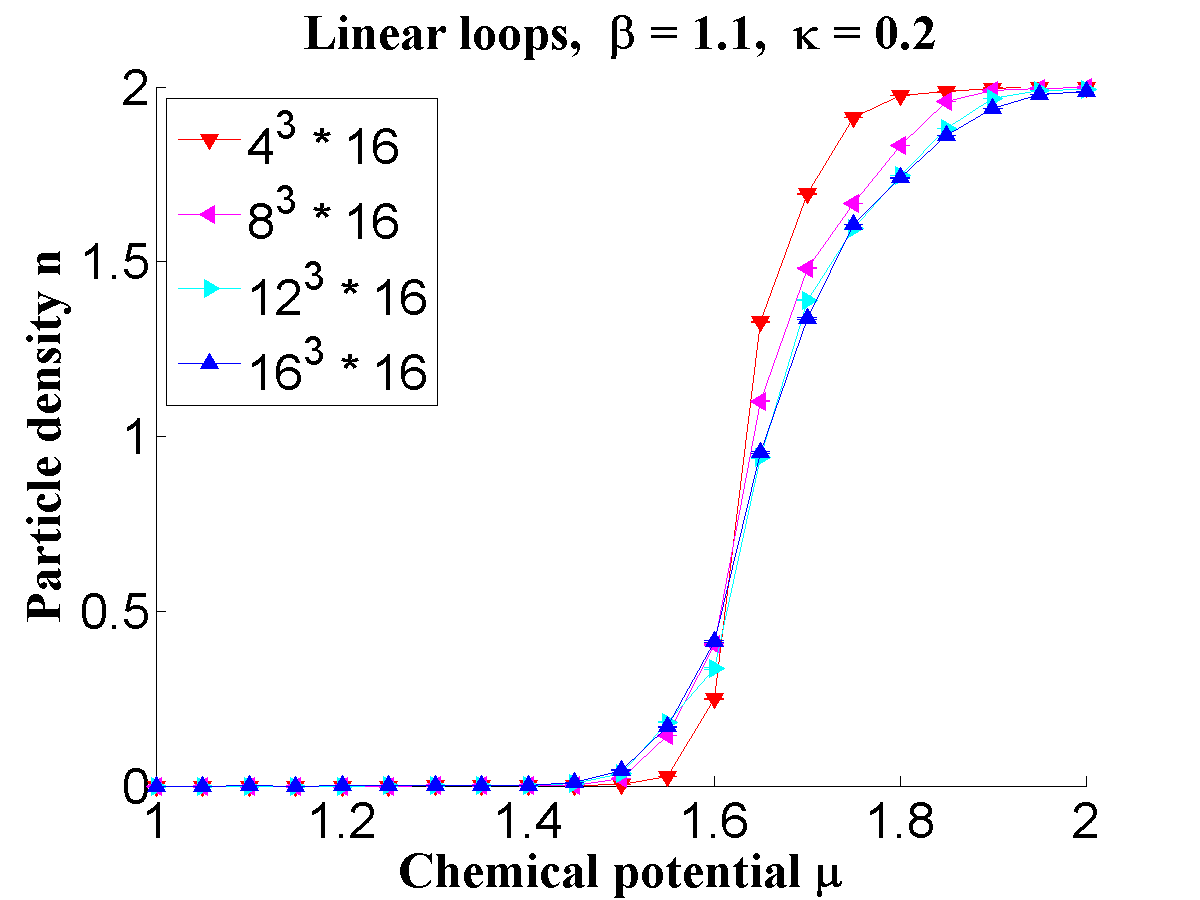}
 
 \vspace{5mm}
 
	\includegraphics[width=0.49\textwidth,clip]{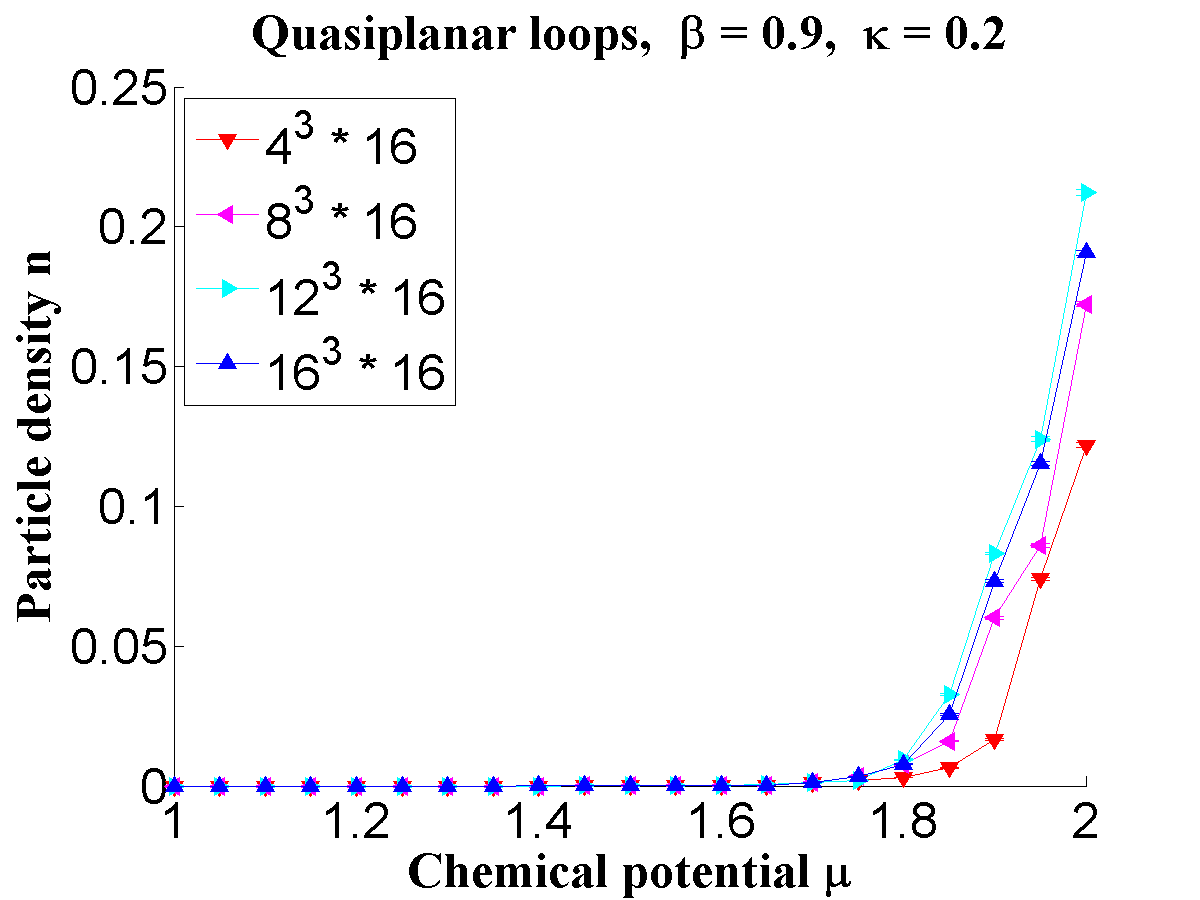}
	\hspace{0.005\textwidth}
	\includegraphics[width=0.49\textwidth,clip]{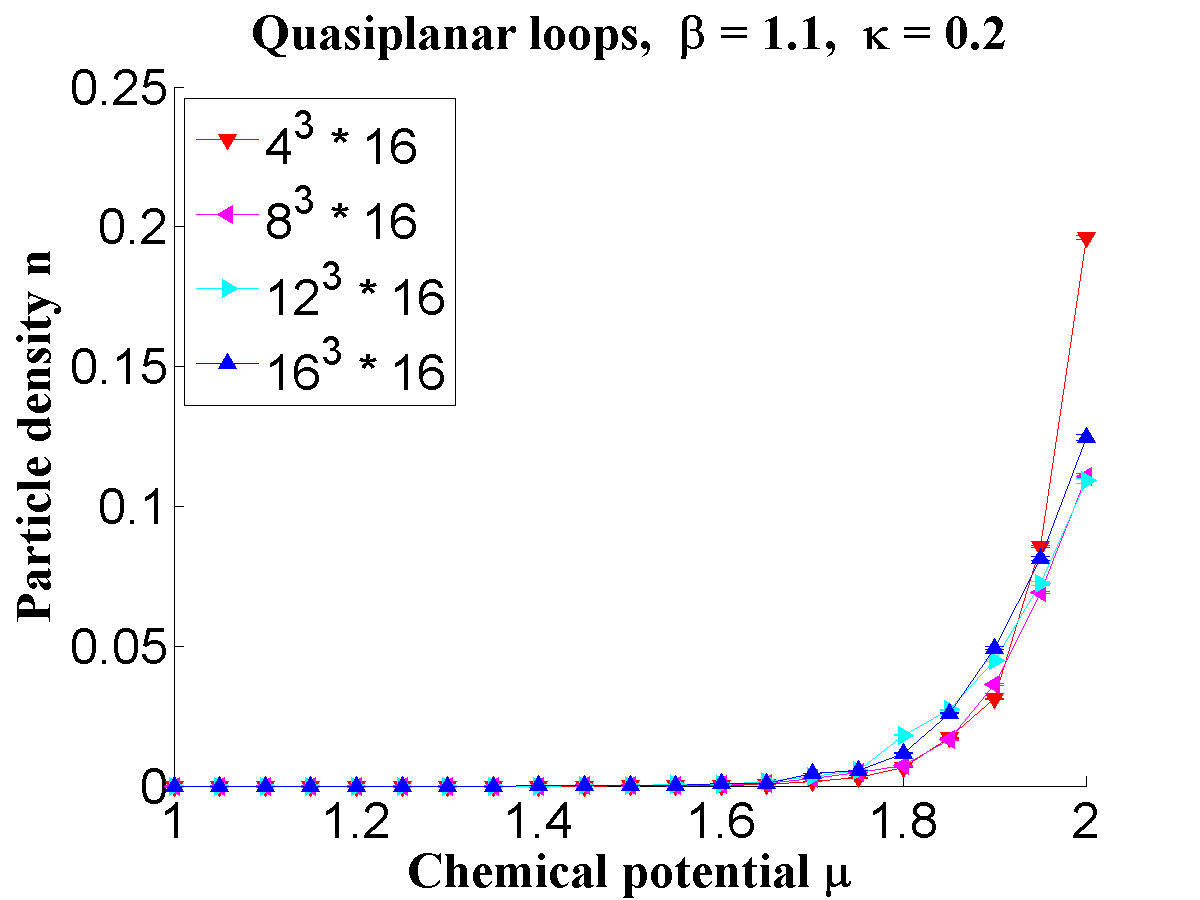}
	\caption{The fermion number density $n$ as a function of the chemical potential. The top row  
	is for the linear loop model, while the bottom is for quasi-planar loops. On the lhs.\ we show the results for $\beta = 0.9$, i.e., 
	the confining  phase, while the rhs.\ is for the Coulomb phase ($\beta = 1.1$). The hopping parameter is set to $\kappa = 0.2$ 
	and we compare the results for different spatial volumes.}
	\label{fig1}
\end{figure}

\begin{figure}[t]
	\centering
	\includegraphics[width=0.48\textwidth,clip]{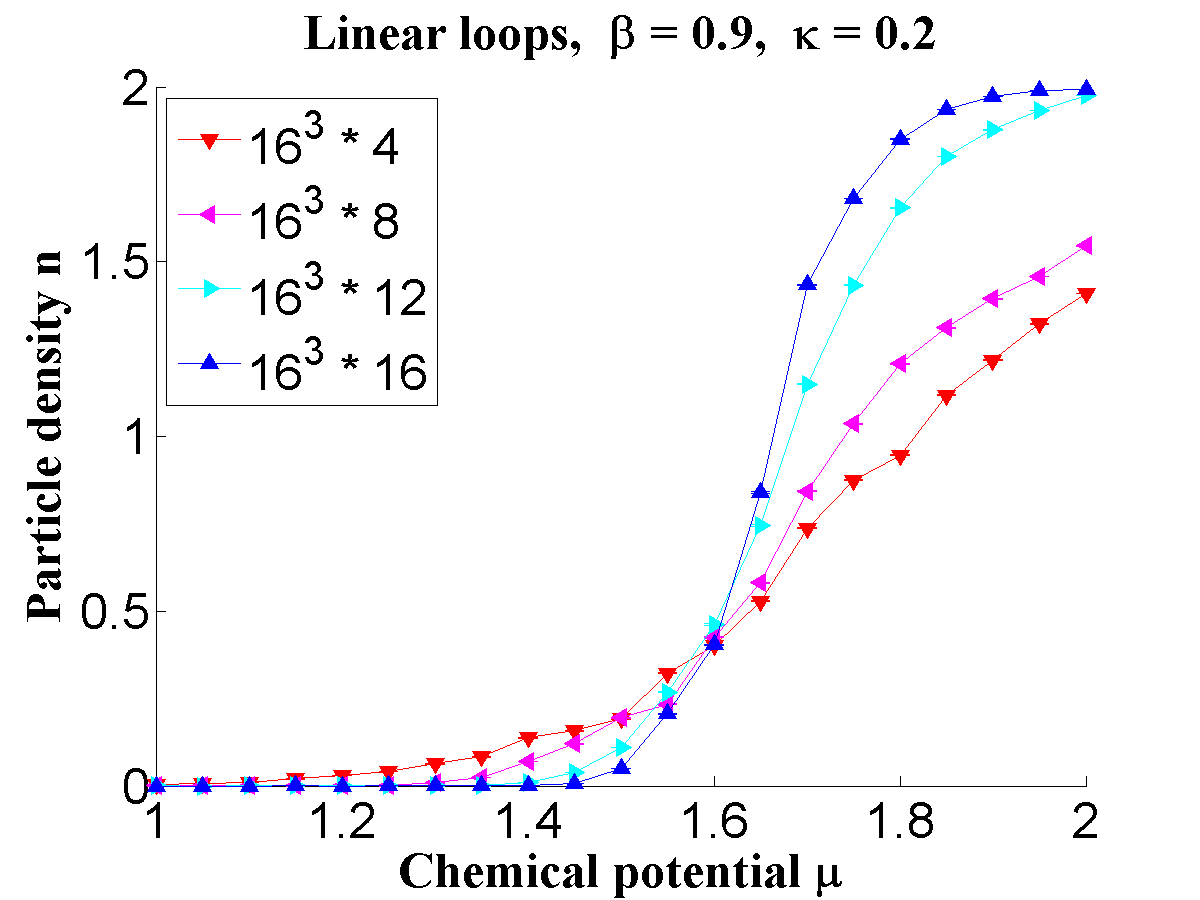}
	\hspace{0.005\textwidth}
	\includegraphics[width=0.48\textwidth,clip]{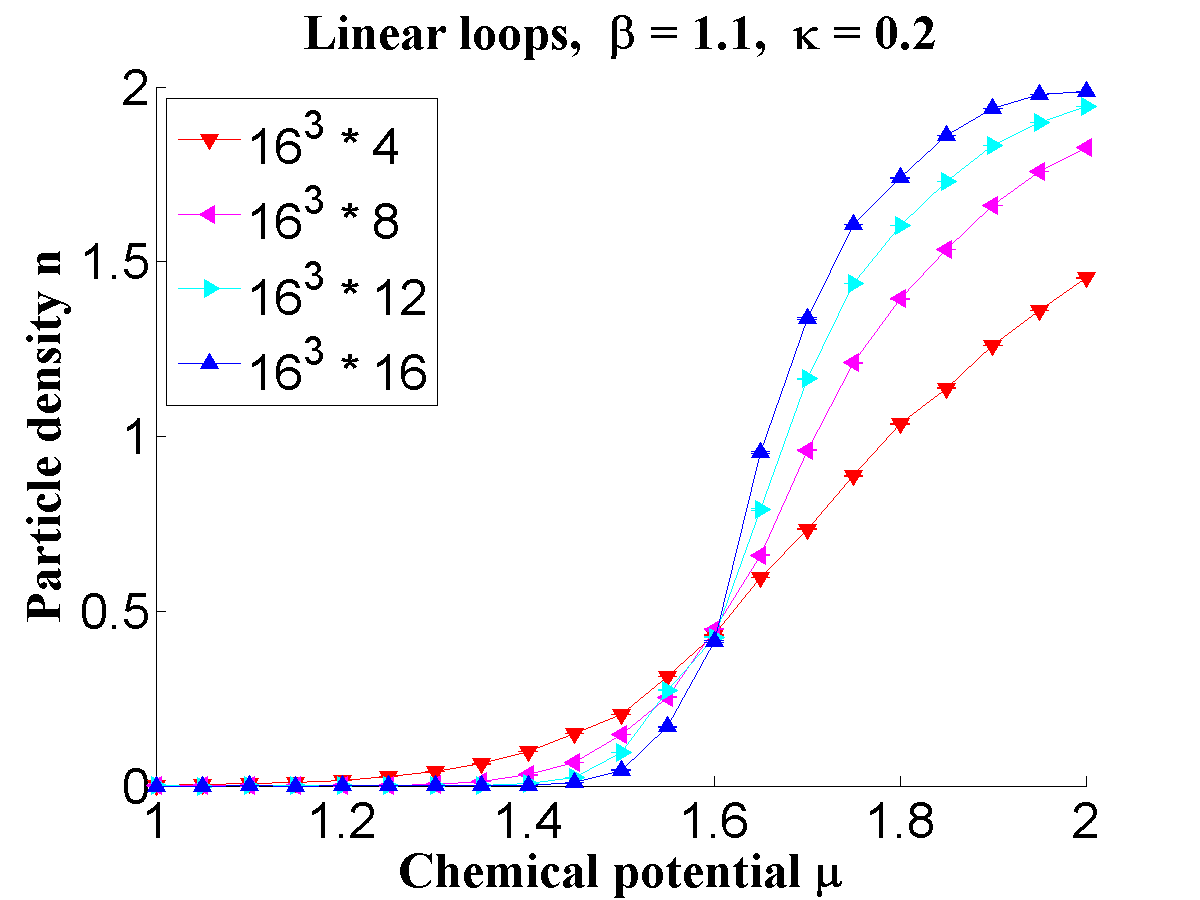}
	\caption{The fermion number density $n$ as a function of the chemical potential for the linear loop model.
	We compare the results for different temperatures $T$, i.e., we vary the temporal extent at a fixed spatial volume of $16^3$. 
	On the lhs.\ we show the results for $\beta = 0.9$, i.e., 
	the confining  phase, while the rhs.\ is for the Coulomb phase ($\beta = 1.1$). The hopping parameter is set to $\kappa = 0.2$ 
	and we compare different $N_T$.}
	\label{fig2}
\end{figure}

\hspace*{10mm}
The simulations were performed on various lattice sizes with $N_S$ and $N_T$ ranging from 4 to 16. We typically used $O(10^4)$
sweeps combining all updates discussed above for equilibration and the configurations used for measuring the observables
are separated by $O(10)$ combined sweeps for decorrelation. The statistics for the observables is typically $O(10^4)$ measurements and
we show statistical errors evaluated with the jackknife method. For the results presented here we used $\kappa = 0.2$ and compared 
the behavior at $\beta = 0.9$ (confined phase) with $\beta = 1.1$ (Coulomb phase).

\hspace*{10mm}
In Fig.~\ref{fig1} we show the results for $n$ as a function of $\mu$ for both the linear loop model (top row of plots) and the quasi-planar model 
(bottom). We compare the results for different spatial volumes to study the emergence of a condensation transition. We observe such a 
transition in both the confining and the Coulomb phases of both models, although at different values of $\mu$. A striking difference is that 
for the linear loop model we clearly observe saturation of the particle density in the condensed phase at a value of $n = 2$, as is expected 
since each loop can be occupied at most twice. For the quasi-planar loops we do not observe saturation. This is due to the fact that there is an 
infinity of quasi-planar loops that can appear in the expansion of the fermion determinant. On the other hand we know that the net particle number
is bounded by $2 N_S^3$, which is the largest power of the fugacity $e^{\mu \beta}$ that can appear in the fermion determinant. This shows that 
subtle cancellations of terms from a hopping expansion of the fermion determinant are an important aspect that should be understood when 
using this expansion for a finite density quantum field theory.  
In Fig.~\ref{fig2} we show for the linear loop model 
the results for the density $n$ as a function of $\mu$ at a fixed spatial volume but now for varying $N_T$, such that
the temperature changes at a fixed physical scale.  As expected for 
a condensation transition, we find that for larger temperature (smaller $N_T$) the transition becomes more washed out. A similar behavior 
was observed for the quasi-planar loops. 

\hspace*{10mm} 
In this exploratory study we analyzed the hopping expansion approach for finding a dual description of fermions a finite chemical potential. 
In particular we were interested in a case where the fermions can be completely dualized, such that effects of the fermion loops can be studied 
without artifacts from, e.g., a strong coupling expansion. We find that the complete dualization of the gauge fields leads to interesting 
cancellations of minus signs due to Gauss' law. Furthermore we demonstrate that the appearance of saturation in the condensed phase 
depends on the truncation used for the hopping expansion.

\end{document}